\documentclass[12pt]{article}
\usepackage{a4wide}
\usepackage{latexsym}
\usepackage{amsmath}
\usepackage{amsfonts}
\usepackage{amscd}
\usepackage{cite}
\usepackage{graphicx}
\usepackage{float}

\usepackage{pslatex}
\usepackage[latin1]{inputenc}
\usepackage[OT2,T1]{fontenc}

\newcommand{\bq}{\begin{eqnarray}}
\newcommand{\eq}{\end{eqnarray}}
\newcommand{\eps}{\varepsilon}

\DeclareSymbolFont{cyrletters}{OT2}{wncyr}{m}{n}
\DeclareMathSymbol{\Sha}{\mathalpha}{cyrletters}{"58}

\begin{document}

\thispagestyle{empty}

\begin{flushright}
  MITP/14-008
\end{flushright}

\vspace{1.5cm}

\begin{center}
  {\Large\bf On the solutions of the scattering equations\\
  }
  \vspace{1cm}
  {\large Stefan Weinzierl\\
\vspace{2mm}
      {\small \em PRISMA Cluster of Excellence, Institut f{\"u}r Physik, }\\
      {\small \em Johannes Gutenberg-Universit{\"a}t Mainz,}\\
      {\small \em D - 55099 Mainz, Germany}\\
  } 
\end{center}

\vspace{2cm}

\begin{abstract}\noindent
  {
This paper addresses the question, whether the solutions of the scattering equations 
in four space-time dimensions can be expressed as rational functions of the momentum twistor variables.
This is the case for $n\le5$ external particles.
For general $n$ there are always two solutions, which are rational functions of the momentum twistor variables.
However, the remaining solutions are in general not rational.
In the case $n=6$ the remaining four solutions can be expressed as algebraic functions.
These four solutions are constructed explicitly in this paper.
   }
\end{abstract}

\vspace*{\fill}

\newpage

\section{Introduction}
\label{sect:intro}

Given a set of $n$ momentum vectors $p_1$, ..., $p_n$ corresponding to a scattering event of $n$ massless particles 
the scattering equations are a set of $n$ equations involving these momentum vectors and
$n$ complex numbers $\sigma_1$, ..., $\sigma_n$ such that
\bq
 \sum\limits_{j=1, j \neq i}^n 
 \;\;
 \frac{2 p_i \cdot p_j}{\sigma_i-\sigma_j} & = & 0,
\eq
for all $i\in\{1,2,...,n\}$.
The scattering equations originate from the work of Cachazo, He and Yuan \cite{Cachazo:2013gna,Cachazo:2013hca}.
Solutions of the scattering equations are $n$-tuples of complex numbers $(\sigma_1,...,\sigma_n)$, which satisfy these equations.
Given a solution we can obtain another solution of the scattering equations by acting with an element $g \in \mathrm{SL}(2,{\mathbb C})$
on the former solution. The precise definition of the group action will be given later on.
Solutions, which are related by a $\mathrm{SL}(2,{\mathbb C})$-transformation are called equivalent solutions.
One is interested in the set of all inequivalent solutions of the scattering equations.
It is known that for $n$ external particles there are $(n-3)!$ inequivalent solutions and this number is independent of the space-time dimension \cite{Cachazo:2013gna}.
Solutions for special kinematical configurations have been discussed in \cite{Kalousios:2013eca}.

The scattering equations may look quite innocent, but the set of inequivalent solutions contains significant information
on the Born $n$-gluon partial amplitude and the $n$-graviton amplitude \cite{Cachazo:2013hca,Dolan:2013isa}.
In fact, both amplitudes may be written in a closed form involving a sum over all inequivalent solutions.

In addition, there is a close relation to the BCJ-decomposition of the amplitudes.
The BCJ-decomposition states, that Born amplitudes in massless gauge theories can 
always be put into a form of a pole expansion, such that the kinematical numerators of this form satisfy
anti-symmetry and Jacobi-like relations, whenever the associated colour factors do \cite{Bern:2008qj}.
These numerators are not unique and the non-uniqueness is attributed to a generalised gauge freedom.
This raises immediately the question, what a convenient generalised gauge choice for the BCJ-numerators would be.
Recently, Monteiro and O'Connell have shown that the inequivalent solutions of the scattering solutions define a canonical 
set of BCJ numerators \cite{Monteiro:2013rya}.

The scattering equations are also closely related to twistor string theories \cite{Mason:2013sva,Berkovits:2013xba,Gomez:2013wza}.

Given the importance of the set of inequivalent solutions of the scattering equations brings us to the question on how to find all inequivalent solutions.
Cachazo, He and Yuan have given in \cite{Cachazo:2013gna} an algorithm to find these solutions numerically.

In this paper we investigate if the scattering equations can be solved analytically, at least for small $n$.
Our main focus is hereby on four space-time dimensions, which is the most relevant case for practical applications.
In particular we address the question if the solutions are rational functions of the input data.
The $n$ momenta $p_1$, ..., $p_n$ are not a particular good parametrisation of the input data. 
The momentum vectors satisfy momentum conservation
and the on-shell conditions, which implies non-trivial relations among them.
In four space-time dimensions
a better parametrisation is given by $n$ momentum twistors $Z_1$, ..., $Z_n$ with no constraints among them.
It is easy to see that in four space-time dimensions and for any $n$ there are always at least two solutions of the scattering equations, 
which are rational functions of the momentum twistor variables.
These two solutions follow from the factorisation of the Lorentz invariants into spinor products: $2 p_i \cdot p_j = \langle p_i p_j \rangle [ p_j p_i ]$.
For $n=3$ and $n=4$ these two solutions are related by a $\mathrm{SL}(2,{\mathbb C})$-transformation, for $n \ge 5$ they are inequivalent (for a generic
non-degenerate kinematical configuration).
The number of inequivalent solutions is $1$ for $n=3,4$ and $2$ for $n=5$.
Therefore, these two solutions exhaust the full set of inequivalent solutions for $n \le 5$.

The problem gets more interesting starting from $n=6$. For six external particles we have $3!=6$ inequivalent solutions.
Taking into account the two rational solutions from above, we have four additional inequivalent solutions.
We will show that these four solutions can be obtained from the zeros of a quartic polynomial.
Being of degree $4$, the roots of this polynomial can still be given in terms of radicals, involving square roots and third roots.
Such a solution would reduce to a rational function, 
if all arguments of the square roots and third roots are squares or third powers, respectively.
However, we will show that in the general case they are not.
Thus, the four additional solutions are algebraic functions of the twistor variables, but not rational functions.

This paper is organised as follows: In section~\ref{sect:setup} we introduce the notation and the set-up.
In section~\ref{sect:rational} we discuss the two rational solutions in the case of four space-time dimensions.
All inequivalent solutions for $n\le6$ are discussed in section~\ref{sect:explicit}.
In particular, the four additional solutions for $n=6$ are constructed in section~\ref{sect:explicit_n=6}.
Finally, section~\ref{sect:conclusions} contains our conclusions.
In an appendix we collected useful information on conformal mappings, spinors and momentum twistors.
 
\section{The scattering equations}
\label{sect:setup}

Let $D$ be the dimension of space-time. 
We are mainly interested in the case $D=4$.
In the following we will always indicate, which statements are valid for general $D$ and which statements hold only for $D=4$.
We denote by ${\mathbb C}M$ the complexified Minkowski space, i.e. a complex vector space of dimension $D$.
The Minkowski metric $g_{\mu\nu}=\mathrm{diag}(1,-1,-1,-1,...)$ is extended by linearity.
We further denote by $\Phi_n$ the (momentum) configuration space of $n$ external massless particles:
\bq
 \Phi_n & = &
 \left\{ \left(p_1,p_2,...,p_n\right) \in \left({\mathbb C} M\right)^n | p_1+p_2+...+p_n=0, p_1^2 = p_2^2 = ... = p_n^2 = 0 \right\}.
\eq
In other words, a set $(p_1, p_2, ..., p_n)$ of $n$ momentum vectors belongs to $\Phi_n$ if this set satisfies momentum conservation
and the mass-shell conditions $p_i^2=0$ for massless particles.
We will use the notation
\bq
 s_{ij} = \left(p_i+p_j\right)^2 = 2 p_i \cdot p_j,
 & &
 s_{ijk} = \left(p_i+p_j+p_k\right)^2,
\eq
where $(p_i+p_j)^2=2p_i \cdot p_j$ holds for massless particles.

We further denote by $\hat{\mathbb C} = {\mathbb C} \cup \{\infty\}$.
The space $\hat{\mathbb C}$ is equivalent to the complex projective space ${\mathbb C}{\mathbb P}^1$.
Let us state the scattering equations:
Given a momentum configuration $(p_1,p_2,...,p_n)\in\Phi_n$ we look for $n$ complex numbers
$(\sigma_1,\sigma_2, ..., \sigma_n) \in \hat{\mathbb C}^n$ such that the $n$ scattering equations 
\bq
\label{scattering_equations}
 \sum\limits_{j=1, j \neq i}^n 
 \;\;
 \frac{2 p_i \cdot p_j}{\sigma_i-\sigma_j} & = & 0,
 \;\;\;\;\;\;\;\;\; i \in \{1,2,...,n\}
\eq
are satisfied.
These equations appeared in the work of Cachazo, He and Yuan \cite{Cachazo:2013gna,Cachazo:2013hca}.
In general, the solution to these equations is not unique.
We are interested in finding all solutions to the scattering equations.
The scattering equations are invariant under $\mathrm{GL}(2,{\mathbb C})$.
Let
\bq
 g = \left(\begin{array}{cc} a & b \\ c & d \\ \end{array} \right) & \in & \mathrm{GL}(2,{\mathbb C}).
\eq
Each $g \in \mathrm{GL}(2,{\mathbb C})$ defines an automorphism of $\hat{\mathbb C}$ as follows:
\bq
 g \cdot \sigma & = &
 \frac{a \sigma + b}{c \sigma + d},
 \;\;\;\;\;\; \sigma \in \hat{\mathbb C}.
\eq
In other words, we have a group homomorphism
\bq
\label{group_homomorphism}
 \mathrm{GL}(2,{\mathbb C}) & \rightarrow & \mathrm{Aut}\left(\hat{\mathbb C}\right),
\eq
where we denoted by $\mathrm{Aut}(\hat{\mathbb C})$ the automorphism group of $\hat{\mathbb C}$.
The kernel in eq.~(\ref{group_homomorphism}) is given by elements of the form
\bq
 g & = & \left(\begin{array}{cc} a & 0 \\ 0 & a \\ \end{array} \right),
 \;\;\;\;\;\; a \in {\mathbb C}^\ast.
\eq
We could therefore in addition impose $\det g = 1$ and restrict our attention to group elements $g\in \mathrm{SL}(2,{\mathbb C})$. 
However, in this paper it will be more convenient to work with $g \in \mathrm{GL}(2,{\mathbb C})$.
We further set
\bq
 g \cdot \left(\sigma_1, \sigma_2, ..., \sigma_n \right)
 & = &
 \left(g \cdot \sigma_1, g \cdot \sigma_2, ..., g \cdot \sigma_n \right).
\eq
If $(\sigma_1,\sigma_2, ..., \sigma_n)$ is a solution of eq.~(\ref{scattering_equations}), then also
$(\sigma_1',\sigma_2', ..., \sigma_n') = g \cdot (\sigma_1,\sigma_2, ..., \sigma_n)$ is a solution.
In order to see this, we note that
\bq
\label{GL2_trafo}
 \frac{1}{\sigma_i'-\sigma_j'}  
 & = &
 \frac{A}{\sigma_i-\sigma_j} + B,
 \;\;\;\;\;\;
 A = \frac{\left(c \sigma_i + d \right)^2}{ad-bc},
 \;\;\;\;\;\;
 B = -\frac{c\left(c \sigma_i +d \right)}{ad-bc}.
\eq
$A$ and $B$ are independent of $\sigma_j$. When eq.~(\ref{GL2_trafo}) is inserted into the scattering equations,
the terms proportional to $A$ vanish due to our assumption that the unprimed values 
$(\sigma_1,\sigma_2, ..., \sigma_n)$ are a solution of the scattering equations, the terms proportional to $B$
vanish due to momentum conservation and the on-shell condition $p_i^2=0$.
We call two solutions which are related by a $\mathrm{GL}(2,{\mathbb C})$-transformation equivalent solutions.

We are in particular interested in the set of all inequivalent solutions of the scattering equations.
As shown in \cite{Cachazo:2013iaa,Cachazo:2013gna}, there are $(n-3)!$ different solutions not related by a $\mathrm{GL}(2,{\mathbb C})$-transformation

We can check if two solutions $(\sigma_1,...,\sigma_n)$ and $(\sigma_1',...,\sigma_n')$ are related 
by a $\mathrm{GL}(2,{\mathbb C})$-trans\-for\-ma\-tion as follows:
We define $g \in \mathrm{GL}(2,{\mathbb C})$ by
\bq
 g \cdot \sigma_{n-2} = 0,
 \;\;\;\;\;\;
 g \cdot \sigma_{n-1} = 1,
 \;\;\;\;\;\;
 g \cdot \sigma_{n} = \infty,
\eq
and
$g' \in \mathrm{GL}(2,{\mathbb C})$ by
\bq
 g' \cdot \sigma_{n-2}' = 0,
 \;\;\;\;\;\;
 g' \cdot \sigma_{n-1}' = 1,
 \;\;\;\;\;\;
 g' \cdot \sigma_{n}' = \infty.
\eq
Then the two sets of solutions are related by a $\mathrm{GL}(2,{\mathbb C})$-transformation if and only if
\bq
 g \cdot \sigma_j & = & g' \cdot \sigma_j',
 \;\;\;\;\;\; \mbox{for all}\;\; j \in \{1,2,...,n-3\}.
\eq
In this case we have
\bq
 \sigma_j' & = & \left(g'\right)^{-1} \cdot g \cdot \sigma_j,
 \;\;\;\;\;\; j \in \{1,2,...,n\}.
\eq
The $n$ scattering equations in eq.~(\ref{scattering_equations}) are not all independent, only 
$(n-3)$ equations are independent.
There are three trivial relations among them, whose origins are as follows:
First of all, if we sum up the left-hand sides of all equations we find
\bq
 \sum\limits_{i=1}^n
 \;\;
 \sum\limits_{j=1, j \neq i}^n 
 \;\;
 \frac{2 p_i \cdot p_j}{\sigma_i-\sigma_j} & = & 0,
\eq
due to the antisymmetry of the denominators $1/(\sigma_i-\sigma_j)$.
Secondly, using momentum conservation, the on-shell conditions and $\sigma_i=(\sigma_i-\sigma_j)+\sigma_j$
we find
\bq
 \sum\limits_{i=1}^n
 \;\;
 \sigma_i
 \sum\limits_{j=1, j \neq i}^n 
 \;\;
 \frac{2 p_i \cdot p_j}{\sigma_i-\sigma_j} 
 & = & 
 -\sum\limits_{i=1}^n
 \;\;
 \sigma_i
 \sum\limits_{j=1, j \neq i}^n 
 \;\;
 \frac{2 p_i \cdot p_j}{\sigma_i-\sigma_j} 
\eq
and hence
\bq
 \sum\limits_{i=1}^n
 \;\;
 \sigma_i
 \sum\limits_{j=1, j \neq i}^n 
 \;\;
 \frac{2 p_i \cdot p_j}{\sigma_i-\sigma_j} 
 & = & 0.
\eq
The third relation comes from the $\mathrm{GL}(2,{\mathbb C})$-invariance.
This is most easily seen by using the freedom of $\mathrm{GL}(2,{\mathbb C})$-transformations to transform
one value (let's say the $i$-th value) to $\sigma_i=\infty$. The $i$-th equations reads then $0=0$.

\section{The two rational solutions}
\label{sect:rational}

In $D=4$ space-time dimensions two solutions of the scattering equations are easily found: The scalar product of massless four-vectors
factorises into spinor products
\bq
 2 p_i \cdot p_j & = & \langle p_i p_j \rangle \left[ p_j p_i \right].
\eq
The spinor products are reviewed in appendix~\ref{appendix:spinor}.
We will need here the following two properties:
First of all, the spinor products are anti-symmetric in the two arguments:
\bq
 \langle p_i p_j \rangle = - \langle p_j p_i \rangle,
 & &
 [ p_i p_j ] = - [ p_j p_i ].
\eq
Secondly, the spinors are only defined up to a scale. We can use this freedom to make one spinor product to be linear in the
momenta.
One possibility, which takes $[p_j p_i]$ to be linear is given by
\bq
 \langle p_i p_j \rangle^{(a)} = 2 \left( \frac{p_j^{\bot\ast}}{p_j^-} - \frac{p_i^{\bot\ast}}{p_i^-} \right),
 & &
 \left[ p_j p_i \right]^{(a)} = \left( p_i^- p_j^\bot - p_j^- p_i^\bot \right).
\eq
The superscript $(a)$ indicates, that this is one possible definition of the spinor products.
If we set now
\bq
\label{solution_a}
 \left(\sigma_1^{(a)}, ..., \sigma_n^{(a)}\right)
 & = &
 \left( \frac{p_1^{\bot\ast}}{p_1^-}, ..., \frac{p_n^{\bot\ast}}{p_n^-} \right),
\eq
the scattering equations reduce to
\bq
 \sum\limits_{j=1, j \neq i}^n 
 \;\;
 \left[ p_j, p_i \right]^{(a)} & = & 0,
 \;\;\;\;\;\;\;\;\; i \in \{1,2,...,n\}.
\eq
Due to momentum conservation and the linearity as well as anti-symmetry of $[p_j p_i]^{(a)}$ these equations are trivially satisfied.
Thus, eq.~(\ref{solution_a}) is a solution to the scattering equations.

A second possibility, which takes $\langle p_i p_j \rangle$ to be linear is given by
\bq
 \langle p_i p_j \rangle^{(b)} = \left( p_i^- p_j^{\bot\ast} - p_j^- p_i^{\bot\ast} \right),
 \;\;\;
 \left[ p_j p_i \right]^{(b)} = 2 \left( \frac{p_j^\bot}{p_j^-} - \frac{p_i^\bot}{p_i^-} \right).
\eq
Setting
\bq
\label{solution_b}
 \left(\sigma_1^{(b)}, ..., \sigma_n^{(b)}\right)
 & = &
 \left( \frac{p_1^{\bot}}{p_1^-}, ..., \frac{p_n^{\bot}}{p_n^-} \right)
\eq
gives a second solution of the scattering equations.
The scattering equations reduce in this case to
\bq
 \sum\limits_{j=1, j \neq i}^n 
 \;\;
 \langle p_i p_j \rangle^{(b)} & = & 0,
 \;\;\;\;\;\;\;\;\; i \in \{1,2,...,n\},
\eq
which are trivially satisfied due to momentum conservation and the linearity and anti-symmetry of $\langle p_i p_j \rangle^{(b)}$.

Using a $\mathrm{GL}(2,{\mathbb C})$-transformation we can bring these solutions into a canonical form where three variables take the values
$0$, $1$ and $\infty$.
This mapping is reviewed in appendix~\ref{appendix:conformal_mapping}.
Choosing $\sigma_{n-2}=0$, $\sigma_{n-1}=1$ and $\sigma_n=\infty$ we obtain
\bq
\label{rational_solutions}
 \Sigma^{(a)}
 & = &
 \left(
 \frac{\langle p_1 p_{n-2} \rangle \langle p_{n-1} p_n \rangle}{\langle p_1 p_n \rangle \langle p_{n-1} p_{n-2} \rangle},
 \frac{\langle p_2 p_{n-2} \rangle \langle p_{n-1} p_n \rangle}{\langle p_2 p_n \rangle \langle p_{n-1} p_{n-2} \rangle},
 ...,
 \frac{\langle p_{n-3} p_{n-2} \rangle \langle p_{n-1} p_n \rangle}{\langle p_{n-3} p_n \rangle \langle p_{n-1} p_{n-2} \rangle},
 0, 1, \infty \right),
 \nonumber \\
 \Sigma^{(b)}
 & = &
 \left(
 \frac{\left[ p_1 p_{n-2} \right] \left[ p_{n-1} p_n \right]}{\left[ p_1 p_n \right] \left[ p_{n-1} p_{n-2} \right]},
 \frac{\left[ p_2 p_{n-2} \right] \left[ p_{n-1} p_n \right]}{\left[ p_2 p_n \right] \left[ p_{n-1} p_{n-2} \right]},
 ...,
 \frac{\left[ p_{n-3} p_{n-2} \right] \left[ p_{n-1} p_n \right]}{\left[ p_{n-3} p_n \right] \left[ p_{n-1} p_{n-2} \right]},
 0, 1, \infty \right).
\eq
Note that any scale factors of the spinors drop out in the ratio, therefore we don't need to specify any superscript for the spinor products.
Eq.~(\ref{rational_solutions}) holds for any definition of the spinor products.
Eq.~(\ref{rational_solutions}) gives two rational solutions for the scattering equations in the case of four space-time dimensions.

It is often convenient to express the kinematical variables through momentum twistors.
Momentum twistors are reviewed in appendix~\ref{appendix:twistors}.
Momentum twistor variables have the advantage that they are not constrained by non-trivial relations.
This is not the case for usual Lorentz invariants $s_{ij}$, where 
momentum conservation and the on-shell conditions lead to non-trivial relations
among them.
A momentum twistor $Z_\alpha$ is a point in ${\mathbb C}{\mathbb P}^3$. In homogeneous coordinates
we write a momentum twistor as
\bq
 Z_\alpha & = & \left( p_A, \mu_{\dot{A}} \right) 
 = \left( |p+\rangle, \langle \mu+ | \right).
\eq
The index $\alpha$ takes the values $\alpha \in \{ 1, 2, \dot{1}, \dot{2} \}$.
We call $p_1$ $p_2$, $\mu_{\dot{1}}$ and $\mu_{\dot{2}}$ ``momentum twistor variables''.
As with momentum vectors, we will not always write the index $\alpha$ explicitly.
A configuration of $n$ momentum twistors $Z_1$, $Z_2$, ..., $Z_n$ is a 
cyclic ordered set $(Z_1, Z_2, ..., Z_n)$.
Such a configuration defines bra- and ket-spinors as follows:
The spinor $p_{iA}=|p_i+\rangle$ is just the one appearing in the first two components of $Z_i$.
The spinor $p_{i\dot{A}}=\langle p_i+|$ is defined by
\bq
\label{def_bra_p_i}
 p_{i\dot{A}}
 =
 - \frac{\left\langle p_i p_{i+1} \right\rangle}{\left\langle p_{i-1} p_{i} \right\rangle\left\langle p_i p_{i+1} \right\rangle}
   \mu_{(i-1) \dot{A}}
 - \frac{\left\langle p_{i+1} p_{i-1} \right\rangle}{\left\langle p_{i-1} p_{i} \right\rangle\left\langle p_i p_{i+1} \right\rangle}
   \mu_{i \dot{A}} 
 - \frac{\left\langle p_{i-1} p_{i} \right\rangle}{\left\langle p_{i-1} p_{i} \right\rangle\left\langle p_i p_{i+1} \right\rangle}
   \mu_{(i+1) \dot{A}}.
\eq
In eq.~(\ref{def_bra_p_i}) all indices are understood modulo $n$.
Note that the definition of $p_{i\dot{A}}$ depends on the cyclic order.
Spinor products are defined as usual through
\bq
 \left\langle p_i p_j \right\rangle = - \eps^{AB} p_{iA} p_{jB},
 & &
 \left[ p_i q_j \right] = \eps^{\dot{A}\dot{B}} p_{i\dot{A}} p_{j\dot{B}}.
\eq
It is clear from eq.~(\ref{def_bra_p_i}) that the spinors $p_{iA}$ and $p_{i\dot{A}}$ are rational functions of the momentum twistor variables.
Hence also the solutions in eq.~(\ref{rational_solutions}) are rational functions of the momentum twistor variables.

\section{Explicit solutions for $n \le 6$}
\label{sect:explicit}

In this section we discuss the explicit solutions in the case where the number of external particles satisfies $n \le 6$.
As the number of $\mathrm{GL}(2,{\mathbb C})$-inequivalent solutions is $(n-3)!$, we have one solution for $n=3$ and $n=4$, two solutions for
$n=5$ and six solutions for $n=6$.
We recall that in four space-time dimensions we have from eq.~(\ref{rational_solutions}) for all $n$ 
two rational solutions of the scattering equations. 
Of particular interest is therefore the case $n=6$, where for the first time the number of solutions exceeds the number of easily known solutions.

\subsection{The case $n=3$}
\label{sect:explicit_n=3}

The case $n=3$ is trivial. Given a momentum configuration $(p_1,p_2,p_3) \in \Phi_3$ it follows from
\bq
 p_1+p_2+p_3=0, 
 \;\;\;\;\;\;
 p_1^2=0,
 \;\;\;\;\;\;
 p_2^2=0,
 \;\;\;\;\;\;
 p_3^2=0,
\eq
that also
\bq
 2 p_1 \cdot p_2 = 0,
 \;\;\;\;\;\;
 2 p_2 \cdot p_3 = 0,
 \;\;\;\;\;\;
 2 p_3 \cdot p_1 = 0.
\eq
Therefore we have three equations $0=0$, which are satisfied by any triple $\Sigma=(\sigma_1,\sigma_2,\sigma_3) \in {\mathbb C}^3$ 
with $\sigma_i \neq \sigma_j$.
Any two solutions $\Sigma'=(\sigma_1',\sigma_2',\sigma_3')$ and $\Sigma=(\sigma_1,\sigma_2,\sigma_3)$ are related by a $\mathrm{GL}(2,{\mathbb C})$-transformation
\bq
 \left(\sigma_1',\sigma_2',\sigma_3'\right) & = & g \cdot \left(\sigma_1,\sigma_2,\sigma_3\right).
\eq
The statements made above are true for a $D$-dimensional space-time. Now let us specialise to four space-time dimensions.
It follows that the two rational solutions of eq.~(\ref{solution_a}) and eq.~(\ref{solution_b}) 
are related by a $\mathrm{GL}(2,{\mathbb C})$-transformation.
This is evident from eq.~(\ref{rational_solutions}), where both solutions have the canonical form $(0,1,\infty)$.

\subsection{The case $n=4$}
\label{sect:explicit_n=4}

Due to the $\mathrm{GL}(2,{\mathbb C})$-invariance we can make the choice $\sigma_2=0$, $\sigma_3=1$ and
$\sigma_4=\infty$.
With this choice the scattering equations reduce to
\begin{alignat}{4}
 && \frac{2p_1\cdot p_2}{\sigma_1} && - \frac{2p_1\cdot p_3}{1-\sigma_1} && = 0,
 \nonumber \\
 -\frac{2p_1\cdot p_2}{\sigma_1} && && - 2p_2\cdot p_3 && = 0,
 \nonumber \\
 \frac{2p_1\cdot p_3}{1-\sigma_1} && + 2p_2\cdot p_3 && && = 0.
\end{alignat}
As already noted in sect.~\ref{sect:setup}, these equations are not independent. Only one equation out of the three equations is independent.
The equations are (uniquely) solved by
\bq
 \sigma_1 & = & - \frac{2p_1\cdot p_2}{2 p_2 \cdot p_3} = - \frac{s}{t},
\eq
where we used the usual definitions $s=(p_1+p_2)^2$ and $t=(p_2+p_3)^2$ of the Mandelstam variables for processes with four external particles.
All other solutions are obtained from
\bq
 \left(\sigma_1,\sigma_2,\sigma_3,\sigma_4\right)
 & = &
 \left(-\frac{s}{t},0,1,\infty\right)
\eq
by a $\mathrm{GL}(2,{\mathbb C})$-transformation.
Now let us consider the case of four space-time dimensions.
Following the argumentation above, the two rational solutions $\Sigma^{(a)}$ and $\Sigma^{(b)}$ should be identical.
The canonical form of $\Sigma^{(a)}$ and $\Sigma^{(b)}$ is
\bq
 \Sigma^{(a)}
 & = &
 \left( \sigma_1^{(a)}, 0, 1, \infty \right)
 =
 \left( 
 \frac{\langle p_1 p_2 \rangle \langle p_3 p_4 \rangle}{\langle p_1 p_4 \rangle \langle p_3 p_2 \rangle},
 0,1,\infty \right),
 \nonumber \\
 \Sigma^{(b)}
 & = &
 \left( \sigma_1^{(b)}, 0, 1, \infty \right)
 =
 \left( 
 \frac{[ p_1 p_2 ] [ p_3 p_4]}{[ p_1 p_4 ] [ p_3 p_2]},
 0,1,\infty \right).
\eq
We have to show
\bq
 \sigma_1^{(a)}
 =
 \sigma_1^{(b)}
 =
 - \frac{s}{t}.
\eq
We start with $\sigma_1^{(a)}$. 
Multiplying numerator and denominator with $[p_4 p_1] [p_2 p_3]$ we obtain
\bq
 \sigma_1^{(a)} & = &
 \frac{\langle p_1 p_2 \rangle [ p_2 p_3 ] \langle p_3 p_4 \rangle [p_4 p_1]}{\left(2 p_2 \cdot p_3\right)\left(2 p_4 \cdot p_1\right)}
 =
 \frac{1}{t^2} \mathrm{Tr}_-\left( p\!\!\!\!/_1 p\!\!\!\!/_2 p\!\!\!\!/_3 p\!\!\!\!/_4 \right),
\eq
with $p\!\!\!\!/ = p^\mu \gamma_\mu$ and $\mathrm{Tr}_-(p\!\!\!\!/...)$ indicates that a projector $\frac{1}{2}(1-\gamma_5)$ is inserted before
$p\!\!\!\!/$.
Evaluating the trace we obtain
\bq
 \mathrm{Tr}_-\left( p\!\!\!\!/_1 p\!\!\!\!/_2 p\!\!\!\!/_3 p\!\!\!\!/_4 \right)
 = 
 2 \left[ \left(p_1 \cdot p_2\right)\left(p_3 \cdot p_4\right)
        - \left(p_1 \cdot p_3\right)\left(p_2 \cdot p_4\right)
        + \left(p_1 \cdot p_4\right)\left(p_2 \cdot p_3\right)
  \right]
 =
 - s t.
 \;\;
\eq
A term proportional to $\eps_{\mu\nu\rho\sigma}p_1^\mu p_2^\nu p_3^\rho p_4^\sigma$ is absent, since the four momenta $p_1$, $p_2$, $p_3$ and $p_4$
are linear dependent.
Thus we find
\bq
 \sigma_1^{(a)} & = & - \frac{s}{t}.
\eq
The argumentation for $\sigma_1^{(b)}$ is similar and we equally find
\bq
 \sigma_1^{(b)} & = &
 \frac{[ p_1 p_2 ] [ p_3 p_4]}{[ p_1 p_4 ] [ p_3 p_2]}
 =
 \frac{1}{t^2} \mathrm{Tr}_+\left( p\!\!\!\!/_1 p\!\!\!\!/_2 p\!\!\!\!/_3 p\!\!\!\!/_4 \right)
 = - \frac{s}{t}.
\eq

\subsection{The case $n=5$}
\label{sect:explicit_n=5}

In this case we make the choice $\sigma_3=0$, $\sigma_4=1$, $\sigma_5=\infty$.
With this choice the scattering equations read
\begin{alignat}{5}
                                 && \frac{s_{12}}{\sigma_1-\sigma_2} && +\frac{s_{13}}{\sigma_1} && -\frac{s_{14}}{1-\sigma_1} && = 0,
 \nonumber \\
 \frac{s_{12}}{\sigma_2-\sigma_1} &&                                 && +\frac{s_{23}}{\sigma_2} && -\frac{s_{24}}{1-\sigma_2} && = 0,
 \nonumber \\
 -\frac{s_{13}}{\sigma_1}         && -\frac{s_{23}}{\sigma_2}         &&                        && -s_{34}                    && = 0,
 \nonumber \\
 \frac{s_{14}}{1-\sigma_1}        && +\frac{s_{24}}{1-\sigma_2}        && +s_{34}                  &&                           && = 0,
\end{alignat}
Only two of these equations are independent and it is convenient to take the third and the fourth equation as the independent equations.
We can use the third equation in order to express $\sigma_2$ in terms of $\sigma_1$:
\bq
\label{equation_sigma_2}
 \sigma_2 & = & - \frac{s_{23} \sigma_1}{s_{34}\sigma_1+s_{13}}.
\eq
Substituting this expression for $\sigma_2$ in the fourth equation yields
a quadratic equation for $\sigma_1$:
\bq
 s_{15} s_{34} \sigma_1^2 
 + \left[ s_{14} s_{35} - s_{13} s_{45} - s_{15} s_{34} \right] \sigma_1
 +s_{13} s_{45}
 & = & 
 0.
\eq
We therefore obtain two solutions for $\sigma_1$, given by
\bq
 \sigma_1^{(1)} = \frac{s_{13} s_{45} - s_{14} s_{35} + s_{15} s_{34} - \sqrt{D}}{2 s_{15} s_{34}},
 & &
 \sigma_1^{(2)} = \frac{s_{13} s_{45} - s_{14} s_{35} + s_{15} s_{34} + \sqrt{D}}{2 s_{15} s_{34}},
\eq
where the discriminant $D$ is given by the $4 \times 4$-Gram determinant
\bq
 D
 & = &
 \mathrm{Gram}\left(p_1,p_3,p_4,p_5\right)
 =
 \left| \begin{array}{cccc}
 0 & s_{13} & s_{14} & s_{15} \\
 s_{13} & 0 & s_{34} & s_{35} \\
 s_{14} & s_{34} & 0 & s_{45} \\
 s_{15} & s_{35} & s_{45} & 0 \\
 \end{array} \right|.
\eq
Each of the two values for $\sigma_1$ determines through eq.~(\ref{equation_sigma_2}) a corresponding value for $\sigma_2$.
We find
\bq
 \sigma_2^{(1)} = \frac{s_{23}s_{45}-s_{24}s_{35}+s_{25}s_{34} + \sqrt{D}}{2 s_{25} s_{34}},
 & &
 \sigma_2^{(2)} = \frac{s_{23}s_{45}-s_{24}s_{35}+s_{25}s_{34} - \sqrt{D}}{2 s_{25} s_{34}}.
\eq
We thus have the two solutions
\bq
\label{solution_n=5_D}
 \Sigma^{(1)} = \left( \sigma_1^{(1)}, \sigma_2^{(1)}, 0, 1, \infty \right),
 & &
 \Sigma^{(2)} = \left( \sigma_1^{(2)}, \sigma_2^{(2)}, 0, 1, \infty \right).
\eq
Eq.(\ref{solution_n=5_D}) is valid in $D$ dimensions.
Now let us specialise to $D=4$ dimensions.
From the spinor products we obtain the two solutions
\bq
 \Sigma^{(a)}
 & = &
 \left( \sigma_1^{(a)}, \sigma_2^{(a)}, 0, 1, \infty \right)
 =
 \left(
 \frac{\langle p_1 p_3 \rangle \langle p_4 p_5 \rangle}{\langle p_1 p_5 \rangle \langle p_4 p_3 \rangle},
 \frac{\langle p_2 p_3 \rangle \langle p_4 p_5 \rangle}{\langle p_2 p_5 \rangle \langle p_4 p_3 \rangle},
 0, 1, \infty \right),
 \nonumber \\
 \Sigma^{(b)}
 & = &
 \left( \sigma_1^{(b)}, \sigma_2^{(b)}, 0, 1, \infty \right)
 =
 \left(
 \frac{\left[ p_1 p_3 \right] \left[ p_4 p_5 \right]}{\left[ p_1 p_5 \right] \left[ p_4 p_3 \right]},
 \frac{\left[ p_2 p_3 \right] \left[ p_4 p_5 \right]}{\left[ p_2 p_5 \right] \left[ p_4 p_3 \right]},
 0, 1, \infty \right).
\eq
The solutions $\Sigma^{(a)}$ and $\Sigma^{(b)}$ are identical to $\Sigma^{(1)}$ and $\Sigma^{(2)}$.
In order to see this, we look at $\sigma_1^{(a)}$. We have
\bq
 \sigma_1^{(a)}
 & = &
 \frac{\langle p_1 p_3 \rangle \langle p_4 p_5 \rangle}{\langle p_1 p_5 \rangle \langle p_4 p_3 \rangle}
 =
 \frac{1}{s_{15}s_{34}} \mathrm{Tr}_-\left( p\!\!\!\!/_1 p\!\!\!\!/_3 p\!\!\!\!/_4 p\!\!\!\!/_5 \right)
  \nonumber \\
 & = &
 \frac{1}{2 s_{15}s_{34}} \left( s_{13} s_{45} - s_{14} s_{35} + s_{15} s_{34} - 4 i \eps(p_1,p_3,p_4,p_5) \right).
\eq
with $\eps(p_1,p_2,p_3,p_4)=\eps_{\mu\nu\rho\sigma}p_1^\mu p_2^\nu p_3^\rho p_4^\sigma$.
We further have
\bq
 \left(4 i \eps(p_1,p_3,p_4,p_5) \right)^2 & = & D.
\eq
Taking $\sqrt{D}=4i\eps(p_1,p_3,p_4,p_5)$ we thus have
\bq
 \Sigma^{(a)} = \Sigma^{(1)},
 & &
 \Sigma^{(b)} = \Sigma^{(2)}.
\eq
Let us further note that in the case where the external momenta $p_1$, ..., $p_5$ are real, the two solutions
$\Sigma^{(1)}$ and $\Sigma^{(2)}$ are related by complex conjugation.

\subsection{The case $n=6$}
\label{sect:explicit_n=6}

In this case we make the choice $\sigma_4=0$, $\sigma_5=1$, $\sigma_6=\infty$.
The scattering equations read then
\begin{alignat}{6}
                                 &&  \frac{s_{12}}{\sigma_1-\sigma_2} && +\frac{s_{13}}{\sigma_1-\sigma_3} && +\frac{s_{14}}{\sigma_1} && -\frac{s_{15}}{1-\sigma_1} && = 0,
 \nonumber \\
 -\frac{s_{12}}{\sigma_1-\sigma_2} &&                                  && +\frac{s_{23}}{\sigma_2-\sigma_3} && +\frac{s_{24}}{\sigma_2} && -\frac{s_{25}}{1-\sigma_2} && = 0,
 \nonumber \\
 -\frac{s_{13}}{\sigma_1-\sigma_3} && -\frac{s_{23}}{\sigma_2-\sigma_3} &&                                  && +\frac{s_{34}}{\sigma_3} && -\frac{s_{35}}{1-\sigma_3} && = 0,
 \nonumber \\
 -\frac{s_{14}}{\sigma_1}         && -\frac{s_{24}}{\sigma_2}          && -\frac{s_{34}}{\sigma_3}          &&                         && -s_{45}                    && = 0,
 \nonumber \\
 \frac{s_{15}}{1-\sigma_1}        && +\frac{s_{25}}{1-\sigma_2}        && +\frac{s_{35}}{1-\sigma_3}        && +s_{45}                  &&                           && = 0.
\end{alignat}
Again, only three out these five equations are independent and it will be convenient to choose the third, the fourth and the fifth equation as the 
independent ones.
We can use the fourth equation to solve for $\sigma_3$:
\bq
\label{eq_z3}
 \sigma_3 & = &
 - \frac{s_{34} \sigma_1 \sigma_2}{s_{24}\sigma_1 + s_{14} \sigma_2 + s_{45} \sigma_1 \sigma_2}.
\eq
We then plug this expression into the third and fifth equations.
The third equations turns into a cubic equation with respect to $\sigma_2$, the fifth equation turns into a quadratic equation
with respect to $\sigma_2$. The latter equation reads
\bq
\label{eq_z2}
 & &
 \left[ s_{45} s_{345} \sigma_1^2 +\left( s_{14} s_{35} - s_{15} s_{34} + \left( s_{14} - s_{34} + s_{25} + s_{56} \right) s_{45} \right) \sigma_1 + s_{14} \left( s_{25} + s_{56} \right)\right] \sigma_2^2 
 \nonumber \\
 & &
 + \left[ \left( s_{24} s_{35} - s_{25} s_{34} + \left( s_{24} - s_{34} + s_{15} + s_{56} \right) s_{45} \right) \sigma_1^2 
          + \left( s_{15} \left(s_{34}-s_{24} \right) + s_{25} \left( s_{34} - s_{14} \right) 
 \right. \right.
 \nonumber \\
 & & 
 \left. \left.
                   - s_{35} \left( s_{14} + s_{24} \right) 
                   - s_{45}  \left( s_{14} + s_{24} - s_{34} + s_{56} \right) \right) \sigma_1 - s_{14} s_{56} \right] \sigma_2
 \nonumber \\
 & &
 + s_{24} \left( s_{15} + s_{56} \right) \sigma_1^2 - s_{24} s_{56} \sigma_1 
 = 0.
\eq
In the next step one solves this quadratic equation for $\sigma_2$ and one inserts the expression for $\sigma_2$ into the third equation.
Note that a quadratic equation has two solutions. Each solution corresponds to a relation between $\sigma_1$ and $\sigma_2$.
The final solution of the scattering equations has to satisfy only one of these two relations, not both.
The third equation contains now only the unknown variable $\sigma_1$. However, the equation involves a square root.
Squaring the equation eliminates the square root, but looses the information on the sign of the square root.
We thus arrive at a polynomial equation for $\sigma_1$. Dividing out several trivial factors, one obtains a polynomial
equation of order $6$ for $\sigma_1$:
\bq
\label{polynomial6}
 A \sigma_1^6 + B \sigma_1^5 + C \sigma_1^4 + D \sigma_1^3 + E \sigma_1^2 + F \sigma_1 + G & = & 0.
\eq
The coefficients are rather lengthy expressions of the kinematical invariants $s_{ij}$. 
For $n=6$ external particles there are $\frac{1}{2}n(n-3)=9$ independent kinematical invariants $s_{ij}$ for general $D$.
The coefficients can be found in a separate file attached to this publication.
Alternatively this file can be obtained by request from the author.
The roots of eq.~(\ref{polynomial6}) are in general distinct and we can label them by $\sigma_1^{(1)}$, $\sigma_1^{(2)}$, ..., $\sigma_1^{(6)}$.
Each root $\sigma_1^{(i)}$ of eq.~(\ref{polynomial6}) will yield -- when inserted into eq.~(\ref{eq_z2}) -- two solutions for $\sigma_2$.
However only one of them will satisfy the original third equation before squaring. 
Once $\sigma_1$ and $\sigma_2$ are known, $\sigma_3$ is uniquely given by eq.~(\ref{eq_z3}).
Thus there are in total six distinct three-tuples $(\sigma_1^{(i)}, \sigma_2^{(i)}, \sigma_3^{(i)})$, as expected.

Let us now specialise to $D=4$ space-time dimensions.
Then there is an additional constraint: All momenta lie in a four-dimensional space and thus the $5 \times 5$ Gram determinant
\bq
\label{Gram5}
 \mathrm{Gram}\left(p_1,p_2,p_3,p_4,p_5\right) & = & 0
\eq
vanishes.
From the spinor products we obtain the two solutions
\bq
\label{rational_n=6}
 \Sigma^{(a)}
 & = &
 \left(
 \frac{\langle p_1 p_4 \rangle \langle p_5 p_6 \rangle}{\langle p_1 p_6 \rangle \langle p_5 p_4 \rangle},
 \frac{\langle p_2 p_4 \rangle \langle p_5 p_6 \rangle}{\langle p_2 p_6 \rangle \langle p_5 p_4 \rangle},
 \frac{\langle p_3 p_4 \rangle \langle p_5 p_6 \rangle}{\langle p_3 p_6 \rangle \langle p_5 p_4 \rangle},
 0, 1, \infty \right),
 \nonumber \\
 \Sigma^{(b)}
 & = &
 \left(
 \frac{\left[ p_1 p_4 \right] \left[ p_5 p_6 \right]}{\left[ p_1 p_6 \right] \left[ p_5 p_4 \right]},
 \frac{\left[ p_2 p_4 \right] \left[ p_5 p_6 \right]}{\left[ p_2 p_6 \right] \left[ p_5 p_4 \right]},
 \frac{\left[ p_3 p_4 \right] \left[ p_5 p_6 \right]}{\left[ p_3 p_6 \right] \left[ p_5 p_4 \right]},
 0, 1, \infty \right).
\eq
This implies that in four space-time dimensions the values
\bq
\label{sigma_a_sigma_b}
 \sigma_1^{(a)} = 
 \frac{\langle p_1 p_4 \rangle \langle p_5 p_6 \rangle}{\langle p_1 p_6 \rangle \langle p_5 p_4 \rangle}
 & \mbox{and} &
 \sigma_1^{(b)} = 
 \frac{\left[ p_1 p_4 \right] \left[ p_5 p_6 \right]}{\left[ p_1 p_6 \right] \left[ p_5 p_4 \right]}
\eq
are roots of eq.~(\ref{polynomial6}).
Since
\bq
\label{factor_4dim}
 \left( \sigma_1 - \sigma_1^{(a)} \right) \left( \sigma_1 - \sigma_1^{(b)} \right)
 & = &
 \sigma_1^2 - \frac{s_{14}s_{56}-s_{15}s_{46}+s_{16}s_{45}}{s_{16}s_{45}} \sigma_1 + \frac{s_{14}s_{56}}{s_{16}s_{45}}
\eq
one might wonder if the right-hand side is a factor of eq.~(\ref{polynomial6}) for general $D$.
This is not the case.
The right-hand side of the above equation is a factor of eq.~(\ref{polynomial6}) only on the sub-space defined by the vanishing of the
Gram determinant in eq.~(\ref{Gram5}).
In other words, carrying out a polynomial division with remainder we find that the remainder vanishes for $D=4$ and is non-zero for $D\neq4$.
However, in four space-time dimensions we may divide out the factor of eq.~(\ref{factor_4dim}) and we can reduce the equation for 
the remaining values of $\sigma_1$
to a quartic polynomial:
\bq
 a \sigma_1^4 + b \sigma_1^3 + c \sigma_1^2 + d \sigma_1 + e & = & 0.
\eq
The coefficients $a$, $b$, ..., $e$ are again rather lengthy expressions in the kinematical invariants $s_{ij}$.
They can be found in the additional file attached to this publication.
Alternatively the file can be obtained by request from the author.
The quartic equation has the solutions
\bq
\label{quartic_equation}
 \sigma_1^{(1)}
 = 
 \frac{1}{4a} \left( -b + \sqrt{D_1} + \sqrt{D_2^-} \right),
 & &
 \sigma_1^{(2)}
 = 
 \frac{1}{4a} \left( -b + \sqrt{D_1} - \sqrt{D_2^-} \right),
 \nonumber \\
 \sigma_1^{(3)}
 = 
 \frac{1}{4a} \left( -b - \sqrt{D_1} + \sqrt{D_2^+} \right),
 & &
 \sigma_1^{(4)}
 = 
 \frac{1}{4a} \left( -b - \sqrt{D_1} - \sqrt{D_2^+} \right),
\eq
with 
\bq
 D_1 & = & b^2 - \frac{8}{3} a c + \frac{2}{3} a R + \frac{8}{3} \left(c^2-3 b d + 12 a e \right) \frac{a}{R},
 \nonumber \\ 
 D_2^\pm & = & 3 b^2 - 8 a c - D_1 \pm 2 \frac{\left( b^3 - 4 a b c + 8 a^2 d \right)}{\sqrt{D_1}},
\eq
and
\bq
 R & = & \left( R_1 + 12 \sqrt{R_2} \right)^{\frac{1}{3}},
 \nonumber \\
 R_1 & = & - 288 a c e - 36 b c d + 108 a d^2 + 108 b^2 e + 8 c^3,
 \nonumber \\
 R_2 & = & 
 240 a b d c^2 e 
 +18 a b^2 d^2 e 
 +576 a^2 b d e^2 
 -54 a b c d^3 
 -54 b^3 c d e
 -432 a^2 c d^2 e 
 -432 a b^2 c e^2  
 \nonumber \\
 & &
 -3 b^2 c^2 d^2 
 +384 a^2 c^2 e^2 
 -48 a c^4 e  
 +12 b^3 d^3
 -768 a^3 e^3 
 +81 a^2 d^4 
 +81 b^4 e^2 
 +12 a c^3 d^2 
 \nonumber \\
 & &
 +12 b^2 c^3 e.
\eq
Together with
\bq
 \sigma_1^{(5)} = \sigma_1^{(a)},
 & &
 \sigma_1^{(6)} = \sigma_1^{(b)}
\eq
from eq.~(\ref{sigma_a_sigma_b}) we thus have all six solutions for $\sigma_1$.
For each $\sigma_1^{(i)}$ the corresponding $\sigma_2^{(i)}$ is obtained from
eq.~(\ref{eq_z2}), for $\sigma_2^{(5)}$ and $\sigma_2^{(6)}$ it is of course simpler to use eq.~(\ref{rational_n=6}).
The value $\sigma_3^{(i)}$ is given by eq.~(\ref{eq_z3}).
Again $\sigma_3^{(5)}$ and $\sigma_3^{(6)}$ can be read off directly from eq.~(\ref{rational_n=6}).

It is now relatively easy to verify that the solutions of the quartic equation~(\ref{quartic_equation})
are not rational functions of the momentum twistor variables.
Choosing for example the momentum twistor configuration
\bq
 \left( \begin{array}{cccc}
 Z_{11} & Z_{12} & Z_{1\dot{1}} & Z_{1\dot{2}} \\
 Z_{21} & Z_{22} & Z_{2\dot{1}} & Z_{2\dot{2}} \\
 Z_{31} & Z_{32} & Z_{3\dot{1}} & Z_{3\dot{2}} \\
 Z_{41} & Z_{42} & Z_{4\dot{1}} & Z_{4\dot{2}} \\
 Z_{51} & Z_{52} & Z_{5\dot{1}} & Z_{5\dot{2}} \\
 Z_{61} & Z_{62} & Z_{6\dot{1}} & Z_{6\dot{2}} \\
\end{array} \right)
 & = &
 \left( \begin{array}{cccc}
 5 & 1 & 7 & 11 \\
 13 & 1 & 17 & 19 \\
 23 & 1 & 29 & 31 \\
 37 & 1 & 41 & 43 \\
 47 & 1 & 53 & 59 \\
 61 & 1 & 67 & 71 \\
 \end{array} \right)
\eq
one easily sees that $\sigma_1^{(1)}$, $\sigma_1^{(2)}$, $\sigma_1^{(3)}$ and $\sigma_1^{(4)}$ are not rational numbers.
We recall that a momentum twistor $Z_i$ is a point in ${\mathbb C}{\mathbb P}^3$, and therefore we may choose $Z_{i 2}=1$ for all $i=1,...,6$.
For all other components we have chosen distinct prime numbers.
We have not included the prime number $3$ in the list of numerical values.
This allows us to exclude a trivial dependence on the irrational number $\sqrt{3}$, related to the third or sixth root of unity.

\section{Conclusions}
\label{sect:conclusions}

In this paper we derived the analytic solutions of the scattering equations in four space-time dimensions for up to six external particles.
The solutions for $n\le 5$ are rational functions of the momentum twistor variables.
The factorisation of the Lorentz scalar product in spinor products implies that there are always for any $n$ at least two solutions,
which are rational functions of the momentum twistor variables 
(for $n=3$ and $n=4$ these two solutions are related by an $\mathrm{SL}(2,{\mathbb C})$-transformation and thus equivalent).
The main result of this paper is the discussion of the case $n=6$.
For six external particles there are $(n-3)!=6$ inequivalent solutions, two of them are rational.
We constructed the analytic expressions for the remaining four solutions and showed that they are in general not rational functions
of the momentum twistor variables.

\begin{appendix}

\section{Conformal mapping}
\label{appendix:conformal_mapping}

In this appendix we review the standard form of a conformal mapping.
An automorphism of $\hat{\mathbb C}$ is uniquely specified by three distinct points of $\hat{\mathbb C}$ and their images.
Let $\sigma_{n-2}$, $\sigma_{n-1}$ and $\sigma_{n}$ be three distinct points of $\hat{\mathbb C}$.
The standard automorphism of $\hat{\mathbb C}$ with
\bq
 g \cdot \sigma_{n-2} = 0,
 \;\;\;\;\;\;
 g \cdot \sigma_{n-1} = 1,
 \;\;\;\;\;\;
 g \cdot \sigma_{n} = \infty,
\eq
is given by
\bq
 g \cdot \sigma & = & \frac{\left(\sigma_{n-1}-\sigma_{n}\right)\left(\sigma-\sigma_{n-2}\right)}{\left(\sigma_{n-1}-\sigma_{n-2}\right)\left(\sigma-\sigma_{n}\right)}.
\eq
The inverse mapping with
\bq
 g^{-1} \cdot 0 = \sigma_{n-2},
 \;\;\;\;\;\;
 g^{-1} \cdot 1 = \sigma_{n-1},
 \;\;\;\;\;\;
 g^{-1} \cdot \infty = \sigma_{n},
\eq
is given by
\bq
 g^{-1} \cdot \sigma'
 & = &
 \frac{\sigma_{n-2}\left(\sigma_{n}-\sigma_{n-1}\right)+\sigma_{n}\left(\sigma_{n-1}-\sigma_{n-2}\right)\sigma'}{\sigma_{n}-\sigma_{n-1}+\left(\sigma_{n-1}-\sigma_{n-2}\right)\sigma'}.
\eq

\section{Spinor products}
\label{appendix:spinor}

We define the light-cone coordinates by
\bq
 p^+=
 \frac{1}{\sqrt{2}}
 \left(p^0+p^3\right), 
 \;\;\;
 p^-=
 \frac{1}{\sqrt{2}}
 \left(p^0-p^3\right), 
 \;\;\; 
 p^\bot=
 \frac{1}{\sqrt{2}}
 \left(p^1+ip^2\right), 
 \;\;\;
 p^{\bot\ast}=
 \frac{1}{\sqrt{2}}
 \left(p^1-ip^2\right).
 \nonumber
\eq
The covariant light-cone coordinates are given by
\bq
 x_+=\frac{1}{\sqrt{2}}\left(x_0+x_3\right), 
 \;\;\;
 x_-=\frac{1}{\sqrt{2}}\left(x_0-x_3\right), 
 \;\;\;
 x_{\bot}=\frac{1}{\sqrt{2}}\left(x_1-ix_2\right), 
 \;\;\;
 x_{\bot\ast}=\frac{1}{\sqrt{2}}\left(x_1+ix_2\right).
\eq
With this definition the Minkowski scalar product is given by
\bq
 p^\mu x_\mu & = & p^+ x_+ + p^- x_- + p^\bot x_{\bot} + p^{\bot\ast} x_{\bot\ast}.
\eq
Associated to a null vector $p^\mu$ we have two Weyl spinors $p_A$ and $p_{\dot{A}}$.
Spinor products are defined by
\bq
 \left\langle p q \right\rangle & = & - \eps^{AB} p_A q_B,
 \nonumber \\
 \left[ pq \right] & = & \eps^{\dot{A}\dot{B}} p_{\dot{A}} q_{\dot{B}}.
\eq
We use the convention that $\eps^{12}=\eps^{\dot{1}\dot{2}}=1$.
The spinors $p_A$ and $p_{\dot{A}}$ are only defined up to a scaling
\bq
 p_A \rightarrow \lambda p_A,
 & &
 p_{\dot{A}} \rightarrow \frac{1}{\lambda} p_{\dot{A}}.
\eq
Keeping the scaling freedom, we define the spinors as
\bq
\label{def_spinors}
p_A = \left| p+ \right\rangle = 
  \frac{\lambda_p 2^{\frac{1}{4}}}{\sqrt{p^-}} 
  \left( \begin{array}{c} p^{\bot^\ast} \\ p^- \end{array} \right),
 & &
p^{\dot{A}} = \left| p- \right\rangle = 
 \frac{2^{\frac{1}{4}}}{\lambda_p\sqrt{p^-}} 
 \left( \begin{array}{c} p^- \\ -p^\bot \end{array} \right),
 \nonumber \\
p_{\dot{A}} = \left\langle p+ \right| = 
 \frac{2^{\frac{1}{4}}}{\lambda_p\sqrt{p^-}} 
 \left( p^\bot, p^- \right),
 & &
p^A =
\left\langle p- \right| = 
 \frac{\lambda_p 2^{\frac{1}{4}}}{\sqrt{p^-}} 
 \left( p^-, -p^{\bot^\ast} \right).
\eq
The spinor products are then
\bq
\label{spinor_products}
 \langle p q \rangle 
 & = & 
 \langle p - | q + \rangle 
 = 
 \frac{\sqrt{2} \lambda_p \lambda_q}{\sqrt{p^- q^-}} \left( p^- q^{\bot\ast} - q^- p^{\bot\ast} \right),
 \nonumber \\
 \left[ q p \right] 
 & = & 
 \langle q + | p - \rangle 
 = 
 \frac{\sqrt{2}}{\lambda_p \lambda_q \sqrt{p^- q^-}} \left( p^- q^\bot - q^- p^\bot \right).
\eq 
We have
\bq
 \langle p q \rangle \left[ q p \right]
 & = & 
 2 p q.
\eq
By a suitable choice of the scale factor $\lambda_p$ we can make one spinor product to be linear in the momenta.
We consider the two specific choices
\bq
\label{choices_spinor_products}
 \lambda_p = \frac{2^{\frac{1}{4}}}{\sqrt{p^-}}
 & : &
 \langle p q \rangle = 2 \left( \frac{q^{\bot\ast}}{q^-} - \frac{p^{\bot\ast}}{p^-} \right),
 \;\;\;
 \left[ q p \right] = \left( p^- q^\bot - q^- p^\bot \right),
 \nonumber \\
 \lambda_p = 2^{-\frac{1}{4}}\sqrt{p^-}
 & : &
 \langle p q \rangle = \left( p^- q^{\bot\ast} - q^- p^{\bot\ast} \right),
 \;\;\;
 \left[ q p \right] = 2 \left( \frac{q^\bot}{q^-} - \frac{p^\bot}{p^-} \right).
\eq
With these choices, one of the spinor products is linear in the momenta. 
The other spinor product is given as the difference of two complex numbers, 
where each of the two complex numbers depends only on one momentum.
Other choices are related to these two cases. For example
\bq
 \lambda_p = \frac{2^{\frac{1}{4}}\sqrt{p^-}}{p^{\bot\ast}}
 & : &
 \langle p q \rangle = 2 \left( \frac{p^-}{p^{\bot\ast}} - \frac{q^-}{q^{\bot\ast}} \right),
 \;\;\;
 \left[ q p \right] = \left( p^{\bot\ast} q^+ - q^{\bot\ast} p^+ \right).
\eq
The complex numbers appearing in $\langle p q \rangle$ are related by the transformation $z\rightarrow -1/z$ to the ones
of the first choice in eq.~(\ref{choices_spinor_products}).

\section{Momentum twistors}
\label{appendix:twistors}

We denote a momentum twistor by
\bq
 Z_\alpha & = & \left( p_A, \mu_{\dot{A}} \right) 
 = \left( |p+\rangle, \langle \mu+ | \right).
\eq
The index $\alpha$ takes the values $\alpha \in \{ 1, 2, \dot{1}, \dot{2} \}$.
The scaling behaviour of a momentum twistor is
\bq
 \left( p_A, \mu_{\dot{A}} \right) & \rightarrow & \left( \lambda p_A, \lambda \mu_{\dot{A}} \right),
\eq
and therefore $Z_\alpha \in {\mathbb C}{\mathbb P}^3$.
As with momentum vectors, we will not always write the index $\alpha$ explicitly.
Let us now consider $n$ momentum twistors $Z_1$, $Z_2$, ..., $Z_n$.
The cyclic ordered set $(Z_1, Z_2, ..., Z_n)$ defines a configuration of $n$ momentum vectors with associated spinors as
follows:
\bq
 \left| p_i + \right\rangle & = & \left| p_i + \right\rangle,
 \nonumber \\
 \left\langle p_i + \right|
 & = &
 - \frac{\left\langle p_i p_{i+1} \right\rangle}{\left\langle p_{i-1} p_{i} \right\rangle\left\langle p_i p_{i+1} \right\rangle}
   \left\langle \mu_{i-1} + \right|
 - \frac{\left\langle p_{i+1} p_{i-1} \right\rangle}{\left\langle p_{i-1} p_{i} \right\rangle\left\langle p_i p_{i+1} \right\rangle}
   \left\langle \mu_{i} + \right|
 - \frac{\left\langle p_{i-1} p_{i} \right\rangle}{\left\langle p_{i-1} p_{i} \right\rangle\left\langle p_i p_{i+1} \right\rangle}
   \left\langle \mu_{i+1} + \right|,
 \nonumber \\
 p_i^\mu & = & \frac{1}{2} \left\langle p_i+ \left| \bar{\sigma}^\mu \right| p_i+ \right\rangle,
\eq
with $\bar{\sigma}^{\mu\dot{A}B}=\left(1,\vec{\sigma}\right)$ and $\vec{\sigma}=(\sigma^1,\sigma^2,\sigma^3)$ are the Pauli matrices.

\end{appendix}

\bibliography{/home/stefanw/notes/biblio}
\bibliographystyle{/home/stefanw/latex-style/h-physrev5}

\end{document}